\documentclass[12pt]{article}
\usepackage{epsfig}
\usepackage{float}
\usepackage{graphicx}
\begin{document}
\rightline{NKU-2015-SF3}
\bigskip

\newcommand{\be}{\begin{equation}}
\newcommand{\ee}{\end{equation}}
\newcommand{\noi}{\noindent}
\newcommand{\ra}{\rightarrow}
\newcommand{\bib}{\bibitem}
\newcommand{\refb}[1]{(\ref{#1})}

\newcommand{\bff}{\begin{figure}}
\newcommand{\eff}{\end{figure}}

\begin{center}
{\Large\bf Black holes in massive gravity: quasinormal modes of Dirac field perturbations}

\end{center}
\hspace{0.4cm}
\begin{center}
Sharmanthie Fernando \footnote{fernando@nku.edu}\\
{\small\it Department of Physics \& Geology}\\
{\small\it Northern Kentucky University}\\
{\small\it Highland Heights}\\
{\small\it Kentucky 41099}\\
{\small\it U.S.A.}\\

\end{center}

\begin{center}
{\bf Abstract}
\end{center}

We have studied  quasinormal modes of  spinor $\frac{1}{2}$, massless Dirac field  perturbations of a black hole in massive gravity.   The parameters of the theory, such as the mass of the black hole, the scalar charge of the black hole, mode number  and the multipole  number   are  varied to observe  how the corresponding quasinormal frequencies change. We have also used  the P$\ddot{o}$schl-Teller approximation to reach analytical values   for the frequencies of quasinormal modes  for comparison with the numerically obtained values.  Comparisons are done with the frequencies of the Schwarzschild black hole.

\hspace{0.7cm}


\section{ Introduction}

Massive gravity is an alternative theory to General Relativity where the graviton, which is a spin two field, acquire a mass. There are several reasons to study massive gravity theories: one  of the reasons being able to explain the acceleration of the universe without introducing  the cosmological constant  or dark energy.  It is speculated that by introducing mass for the graviton, the gravity can be modified at the infrared in a such a way as to produce the acceleration of the universe.

The first attempt to include mass for the graviton was done by Fierz and Pauli in 1939 \cite{pauli}. In comparison with GR where the graviton has two degrees of freedom, in a massive gravity theory the graviton has five degrees of freedom.

During the last few years, significant progress has been done on massive gravity theories. For example, a particular massive gravity theory called dRGT theory  \cite{drgt1}\cite{drgt2}  is found to be free from Boulware-Deser ghosts. Two other well known massive gravity theories which are also know to be free from ghosts are,  DGP model \cite{dgp} and the ``new massive gravity theory''   in three dimensions \cite{town}. There are many interesting works related to massive gravity in the literature and we find it difficult to discuss all; we will direct the reader to  excellent reviews on the topic by de Rham \cite{claudia} and  Hinterbichler \cite{kurt} instead.

In this paper, we focus on a massive gravity theory which is a Lorentz violating theory. In this model, Lorentz symmetry is broken spontaneously by four scalar fields called Goldstone bosons. These field are coupled to gravity through derivative coupling. When Lorentz symmetry is broken spontaneously, the graviton acquire a mass very similar to the Higgs mechanism.  A nice review of Lorentz violating massive gravity theory can be found in \cite{dubo} \cite{ruba2}.

The Lorentz violating theory of massive gravity considered in this paper is described by the following action:

\be \label{action}
S = \int d^4 x \sqrt{-g } \left[ \frac{R}{ 16 \pi}  + \Lambda^4 \mathcal{F}( X, W^{ij}) \right]
\ee
Here $R$ is the scalar curvature of the space-time geometry and $\mathcal{F}$  is a function of $X$ and $W^{i j}$. The functions $X$ and $W^{i j}$ are functions of scalar fields $\Phi^{0}, \Phi^{i}$ and are defined as,
\be
X = \frac{\partial^{\mu} \Phi^0 \partial_{\mu} \Phi^0 }{ \Lambda^4}
\ee
\be
W^{i j} = \frac{\partial^{\mu} \Phi^i \partial_{\mu} \Phi^j}{ \Lambda^4} - \frac{\partial^{\mu} \Phi^i \partial_{\mu} \Phi^0  \partial^{\nu} \Phi^j \partial_{\nu} \Phi^0 }{ \Lambda^4 X}
\ee
The theory described by action in eq.$\refb{action}$ is the low-energy effective theory below the ultra-violet cutoff $\Lambda$ and a perturbative analysis on the theory calculate the value of $\Lambda$ to be in the order of $ \sqrt{ m M_{pl}}$ where $m$  is the graviton mass and $M_{pl}$ the Plank mass\cite{luty}  \cite{dubo} \cite{ruba} \cite{pilo1}. The scalar fields $\Phi^0, \Phi^i$  are responsible for spontaneously breaking Lorentz symmetry. When  symmetry is broken, the scalar fields $\Phi^0, \Phi^i$ acquire a vacuum expectation value and  are called Goldstone fields.

Studies of perturbations and quasi normal modes (QNM) of black holes have a rich history. The founders of black hole perturbations were Regee and Wheeler where they established the equations of axial perturbations of the Schwarzschild black hole in 1957 \cite{regee}. Thereafter, the contributions from Zerilli, Vishveshwara, Press and Teukolsky put a firm foundation on the area of research on black hole perturbations as an important field of research in gravitational physics. A nice review on black hole perturbations and current methods used can be found in \cite{pani}.
QNM are a set of modes arising during the intermediate stage of a black hole perturbation. The frequencies of QNM are complex and only depend on the properties of the black hole such as the mass, spin and the charge. Studies of QNM frequencies has attracted great  attention due to variety reasons: one of the prominent being the current attempts to observe gravitational waves. If current gravitational wave detectors can detect signals emitted from black holes, then we will be able to specify their properties of such objects in the universe \cite{ferra}.  An exhaustive reasons as to why we need to study QNM frequencies are given in the nice review by Konoplya and Zhidenko \cite{kono1}.

In this paper we focus on the spin $\frac{1}{2}$, massless  Dirac spinor field perturbation of a black hole in massive gravity. There are several works which have focused on the QNM of Dirac perturbations of black holes:  Dirac perturbations of the Lifshitz black hole in 4 dimensions were studied in \cite{cata3}.  Scalar and Dirac perturbations of the Kerr-Newmann-de Siter black hole were analyzed by Konoplya in \cite{kono7}.  Spinor perturbations of regular black holes where nonlinear electrodynamics are coupled to gravity were studied by Li and Ma in \cite{ma}. Decay of spinor half field around a Born-Infeld black hole was studied by Fernando \cite{fernando6}.

The paper is organized as follows: in section 2, the black hole in massive gravity is introduced. In section 3, the equations for the massless Dirac field of spin $\frac{1}{2}$ is developed and presented. In section 4, the WKB approach is employed to compute QNM frequencies by varying the parameters in the theory.  The P$\ddot{o}$schl-Teller approximation is used to find analytical values for $\omega$ in section 5. Finally in section 6 the conclusion is given.

%%%%%%%%%%%%%%%%%%%

\section{ Black holes in massive gravity}

In this section we will present the basic characteristics of the black hole in massive gravity considered in this paper. A detailed derivation can be found in \cite{tinya} and \cite{pilo}.

The geometry is described by the metric  given by,
\begin{equation} \label{metric}
ds^2 = - f(r) dt^2 + \frac{ dr^2}{ f(r)} + r^2 ( d \theta^2 + sin^2 \theta d \phi^2)
\end{equation}
where,
\begin{equation}
f(r) = 1 - \frac{ 2 M} { r} - \frac{ Q}{r^{\lambda}}
\end{equation}
The scalar fields are given by,
\be
\Phi^0 = \Lambda^2 \left( t + z(r) \right); \hspace{1 cm} \Phi^i =   \Lambda^2 x^i
\ee
where
\be
z(r) = \pm \int \frac{ dr} { f(r)} \left[ 1 - f(r) \left( \frac{ Q \lambda(\lambda-1)}{ 12 m^2}\frac{1}{ r^{\lambda+2}} + 1 \right)^{-1} \right]^{1/2}
\ee
Here $m$ is the mass of the graviton and $\lambda$  is a positive constant. $\lambda$  has to be greater than 1 in order  for the solutions to be asymptotically flat and $M$ to be the ADM mass.  $Q$ is a scalar charge and represents a gravity theory with a massive graviton.\\
\noi
The function $\mathcal{F}$ for this particular black hole solution is given by,
\be
\mathcal{F} = \frac{ 12} { \lambda} \left( \frac{ 1}{ X} + p_1 \right) - \left( p_1^3 - 3 p_1 p_2 - 6 p_1 + 2 p_3 - 12\right)
\ee
where,
\be
p_n = Tr( W^n)
\ee

\begin{figure} [H]
\begin{center}
\includegraphics{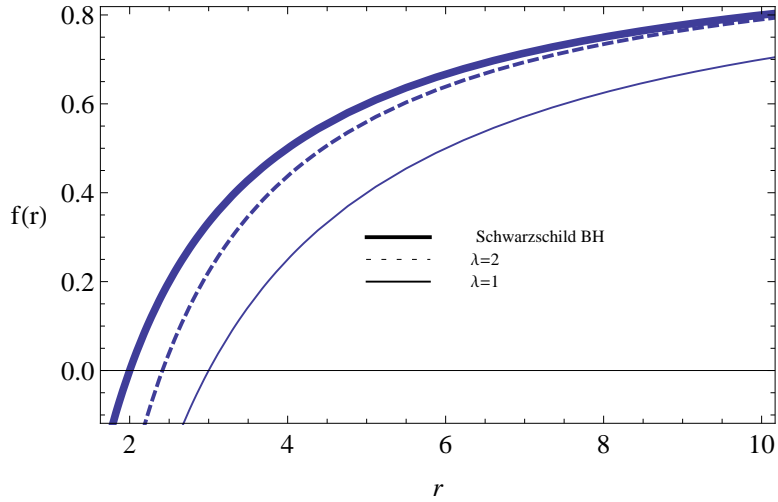}
\caption{The figure shows  $f(r)$ vs $r$. Here $M = 1$ and $Q = 1$.}
\label{frlambda}
 \end{center}
 \end{figure}

There are two possibilities for $Q$ when $M >0$. When $Q > 0$, the geometry is very similar to the Schwarzschild black hole with a single horizon. The function $f(r)$ for this case is given in Fig.$\refb{frlambda}$. The horizon radius for the black hole in massive gravity is larger than the one for the Schwarzschild black hole. When $\lambda \ra \infty$,  $ r_h \ra 2 M$ which is the value for the Schwarzschild black hole.

\begin{figure} [H]
\begin{center}
\includegraphics{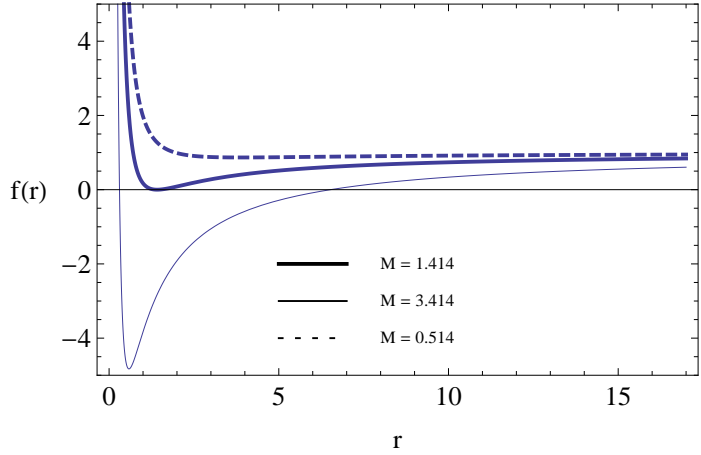}
\caption{The figure shows  $f(r)$ vs $r$. Here $Q = -2$ and $\lambda =2$.}
\label{frq}
 \end{center}
 \end{figure}

When $Q <0$, the geometry is similar to the well known Reissner-Nordstrom charged black hole. There could be two, one or no horizons depending on the parameter of the theory.  The function $f(r)$ for this case is given in Fig.$\refb{frq}$.

When the mass of the black hole $M$ is,
\be
M_{critical} = \frac{ \lambda |Q|^{1/\lambda}}{ 2} \left( \frac{ 1}{ \lambda -1} \right)^{\frac{ \lambda -1}{ \lambda}} 
\ee
the horizons merge. For $M > M_{critical}$,  there will be two horizons. For $M < M_{critical}$, there wont be any horizons and there will be a naked singularity.

\begin{figure} [H]
\begin{center}
\includegraphics{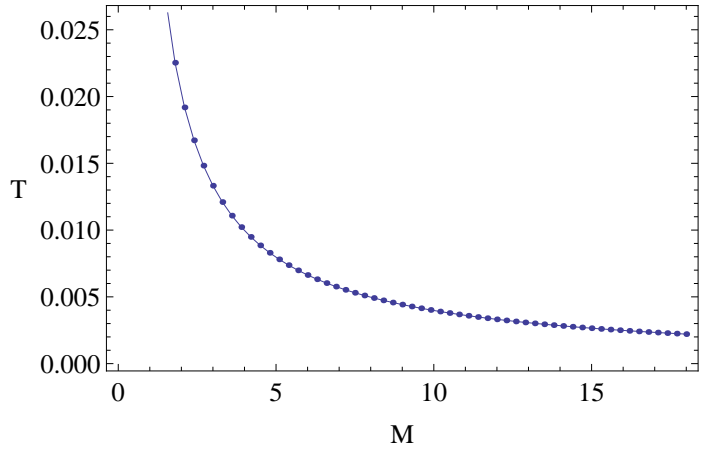}
\caption{The figure shows  $Temperature$ vs $M$. Here $Q = 1$ and $\lambda = 3$.}
\label{tempM}
 \end{center}
 \end{figure}

\begin{figure} [H]
\begin{center}
\includegraphics{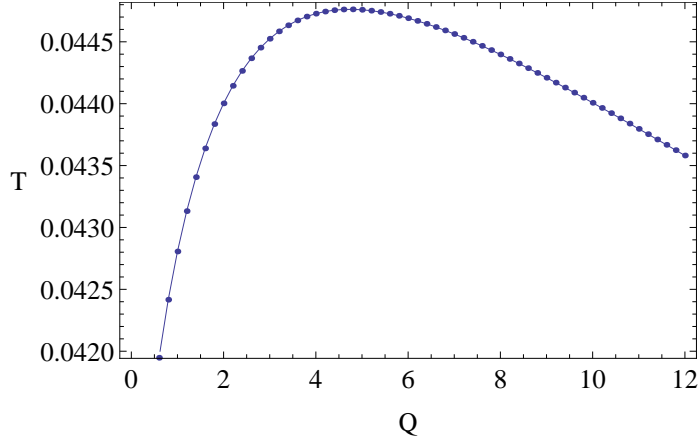}
\caption{The figure shows  $Temperature$ vs $Q$. Here $M = 1$ and $\lambda = 3$.}
\label{tempQ}
 \end{center}
 \end{figure}

The Hawking temperature of the black hole is given by,
 \be
 T_H =  \frac{ 1}{ 4 \pi}  \left| \frac{ df(r)}{ dr} \right|_ { r = r_h} = \frac{1}{ 4 \pi} \left(\frac{ 2 M}{ r_h^2} + \frac{ Q \lambda}{ r_h^{ \lambda + 1}} \right)
 \ee
The temperature is plotted varying the mass $M$ and $Q$ in Fig.$\refb{tempM}$ and Fig.$\refb{tempQ}$. When  M increases, the temperature decreases. This behavior  is similar to the behavior of the temperature of the Schwarzschild black hole.  When Q is increased, the temperature increases to a maximum  and then decreases. Thermodynamics and phase structure of the massive gravity black hole analyzed  in this paper was addressed by Capela and Nardini 
\cite{nar}.

%%%%%%%%%%%%%%%%%%%

\section{A massless, spin $ \frac{1}{2}$  Dirac field around a black hole}

In this section, the Dirac equation around the black hole considered in this paper is presented. The equation for a Dirac spinor field of spin $\frac{1}{2}$ in a curved back ground is given by,
\begin{equation} \label{dirac}
\gamma^a e^{\nu}_{a} ( \partial_{\nu} - \Gamma_{\nu} ) \xi =0
\end{equation}
Here, $e_{\nu}^{a}$ is the tetrads and $e^{\mu}_{a}$ is the inverse of the tetrads. The metric tensor $g_{\alpha \beta}$ of the space-time considered and the tetrads $e^{a}_{\alpha}$ are related by,
\begin{equation}
g_{\alpha \beta} = \eta_{a b} e^{a}_{\alpha} e^{b}_{\beta}
\end{equation}
Here, $\eta_{a b} = ( -1,1,1,1)$  is  the metric of the flat space. The $\gamma^a$ matrices appearing in  eq.$\refb{dirac}$ are  defined by,
\begin{eqnarray}
\gamma^{0} =  \left(\begin{array}{cc} i & 0 \\ 0 & - i
\end{array} \right) ; \;\;\;\; 
\gamma^{a} =  \left(\begin{array}{cc}  0  & i  \sigma^a \\  - i  \sigma^a & 0
\end{array} \right)
\end{eqnarray}
Here, $ \sigma^{a}$  are the well known  Pauli matrices  given by,
\begin{eqnarray}
\sigma^{1} =  \left(\begin{array}{cc} 1 & 0 \\ 0 & -1
\end{array} \right) ; \;\;\;\; 
\sigma^{2} =  \left(\begin{array}{cc}  0 & - i \\ i & 0
\end{array} \right);  \;\;\;\; 
\sigma^{3} = \left(\begin{array}{cc} 0 & 1 \\ 1 & 0
\end{array} \right)
\end{eqnarray}
The $\gamma^a$ matrices satisfy the anti-commuting relations,
\begin{equation}
\{ \gamma^a, \gamma^b \} = 2 \eta^{a b} I
\end{equation}
The spin connections $\Gamma_{\mu}$ in eq.$\refb{dirac}$ are defined as,

\begin{equation}
\Gamma_{\mu} = - \frac{1}{8} [ \gamma^a , \gamma^b ] e^{\nu}_{a} e_{b \nu; \mu}
\end{equation}
with,
 \begin{equation}
e_{ b \nu ; \mu} = \partial_{\mu} e_{b \nu} - \Gamma_{\mu \nu}^{\beta} e_{b \beta}
\end{equation}
$\Gamma_{ \nu \mu} ^{\beta} $ are the Christoffel symbols.

For the metric given in this paper, the  tetrads  are given by,
\begin{equation}
e^a_{\mu} = diag \left( \sqrt{f}, \frac{1}{\sqrt{f}}, r, r sin \theta \right)
\end{equation}
Therefore, the  spin connections for this space-time is given by,

$$ \Gamma_t = \frac{ f'}{ 4 } \gamma^ 0 \gamma^ 1$$
$$ \Gamma_r = 0$$
$$ \Gamma_{\theta} = \frac{ \sqrt{f}}{2 } \gamma^1 \gamma^2$$
\begin{equation}
\Gamma_{\varphi} = \frac{cos \theta }{2} \gamma^2 \gamma^3 + \frac{\sqrt{f}}{2} sin \theta \gamma^1 \gamma^3
\end{equation}

To facilitate computations, the function $\xi$ can be  redefined as,
\begin{equation}
\xi = \frac{\Phi}{ f^{1/4} }
\end{equation}
Then the Dirac equation $\refb{dirac}$ simplifies to,
\begin{equation} \label{dirac2}
\frac{\gamma^0}{\sqrt{f}} \frac{\partial \Phi}{\partial t} + \sqrt{f} \gamma^1 
\left( \frac{\partial}{\partial r} + \frac{1}{r} \right) \Phi +
\frac{\gamma^2}{r} \left( \frac{\partial}{\partial \theta} + \frac{ cot \theta} {2} \right) \Phi + \frac{ \gamma^3} { r sin \theta} 
\frac{ \partial \Phi} { \partial \varphi} = 0
\end{equation}
Before proceeding further, some clarifications are required. Since the field considered here is massless and spin $\frac{1}{2}$, the allowed solutions to the Dirac equation are circularly polarized. More details of this argument can be found in the paper by Brill and Wheeler \cite{brill}. According to Brill and Wheeler \cite{brill},  the spinors considered have right handed circular polarization and the allowable spin states satisfy the identity,
\begin{equation} \label{circular}
( 1 - i \gamma_5 ) \Phi = 0
\end{equation}
Here $\gamma_5$ is given by,
\begin{equation} 
\gamma^5 = \gamma^0 \gamma^1 \gamma^2 \gamma^3
\end{equation}
The eq.$\refb{circular}$ gives a simplified set of components for $\Phi$ as,
\begin{eqnarray}
\Phi =  \left(\begin{array}{l} \Phi_1( t,r,\theta, \varphi) \\ 
\Phi_2( t,r,\theta, \varphi) \\ 
\Phi_1 ( t,r,\theta, \varphi) \\ 
\Phi_2 ( t,r,\theta, \varphi)
\end{array} \right) 
\end{eqnarray}
Hence the Dirac equation yield two identical set of equations each coupling $\Phi_1$ and $\Phi_2$.  The  components  $\Phi_1$ and $\Phi_2$ are redefined as follows:
\begin{eqnarray} \label{new}
\left(\begin{array}{l} \Phi_1( t,r,\theta, \varphi) \\ 
\Phi_2( t,r,\theta, \varphi) 
\end{array} \right) = \left(\begin{array}{l}   \frac{i K(r)}{r} \alpha_1(\theta, \varphi) \\ 
 \frac{G(r)}{r} \alpha_2(\theta, \varphi) 
\end{array} \right) e^{- i \omega t}
\end{eqnarray}
Substituting  eq.$\refb{new}$ to the Dirac equation$\refb{dirac2}$ yields,
\begin{equation} \label{wave1}
\left( \frac{ i \omega r}{ \sqrt{f} } K -  r \sqrt{f} \frac{ dK }{ dr}\right) \frac{1}{G} + \left( \frac{cot \theta}{2} \alpha_2 + \frac{i}{sin \theta} \frac{ \partial \alpha_2} {\partial \varphi} + \frac{\partial \alpha_2} {\partial \theta} \right) \frac{1}{ \alpha_1}=0
\end{equation}

\begin{equation} \label{wave2}
\left( \frac{  i \omega r}{ \sqrt{f} } G +  r \sqrt{f} \frac{ dG}{ dr} \right) \frac{1}{K} + \left( \frac{cot \theta}{2} \alpha_1 - \frac{i}{sin \theta} \frac{ \partial \alpha_1} {\partial \varphi} + \frac{\partial \alpha_1} {\partial \theta} \right) \frac{1}{ \alpha_2}=0
\end{equation}
The angular part of the above equation can be solved in terms of spin-weighted spherical harmonics, $_{s}Y_{lm} $. A lengthy description of spin weighted spherical harmonics can be found in \cite{gold} and \cite{torres}. One can define two operators $\partial_+$ and $\partial_-$ as,
\begin{equation} \label{new1}
\partial_{+} = - \left( \frac{cot \theta}{2}  + \frac{i}{sin \theta} \frac{ \partial } {\partial \varphi} + \frac{\partial} {\partial \theta} \right) 
\end{equation}

\begin{equation} \label{new2}
\partial_{-} = - \left( \frac{cot \theta}{2}  - \frac{i}{sin \theta} \frac{ \partial } {\partial \varphi} + \frac{\partial} {\partial \theta} \right) 
\end{equation}
It was shown  in \cite{gold}  \cite{torres} that the above operators act on spin weighted spherical harmonics $_sY_{lm}$ as ladder operators. In particular if the spin $ s = \frac{1}{2}$,  the above operators gives the following relations;
\begin{equation} \label{spherical}
\partial_+ ( _{-\frac{1}{2}}Y_{lm} ) =  ( l + \frac{1}{2} )  _{\frac{1}{2}}Y_{lm} 
\end{equation}
\begin{equation}
\partial_- ( _{\frac{1}{2}}Y_{lm} ) = - ( l + \frac{1}{2} )  _{-\frac{1}{2}}Y_{lm} 
\end{equation}
Note that $ l  \leq  |s|$ and $l$ has to be half integer. Hence $ l$ can be written in terms of $\mu$ which is a positive  integer as, $ l = \mu - \frac{1}{2}$. Here $\mu$ will be called the multipole number in the rest of the paper which can take any positive integer.

If the functions  $\alpha_1(\theta, \varphi)$ and $\alpha_2(\theta, \varphi)$ are chosen as spin weighted spherical harmonics given by,
\begin{equation}
\alpha_1  =  { _{\frac{1}{2}}Y_{lm}}
\end{equation}
\begin{equation}
\alpha_2  =  {_{-\frac{1}{2}}Y_{lm}}
\end{equation}
the eq.$\refb{wave1}$ and eq.$\refb{wave2}$ can be simplified to be,
\be \label{fun1}
\frac{dG}{dr^*} - i \omega G + W(r)  K = 0
\ee
\begin{equation} \label{fun2}
\frac{dK}{dr^*} + i \omega K + W(r)  G = 0
\end{equation} \label{tor}
$r_*$ is the  ``tortoise'' coordinate given by,
\begin{equation} \label{tortoise}
dr_{*} = \frac{dr}{f}
\end{equation}
and the function $W(r)$ is,
\begin{equation}
W(r) = \frac{\mu \sqrt{f}}{r}
\end{equation}
Two new functions $\beta_{\pm}$ can be  defined as,
\begin{equation}
\beta_{\pm} = K \pm G
\end{equation}
and  eq.$\refb{fun1}$ and eq.$\refb{fun2}$  can be decoupled as,
\begin{equation} \label{final}
\frac{d^2 \beta_{\pm}}{dr^{*2}} + ( \omega^2 - V_{Dirac}^{\pm} ) \beta_{\pm} = 0
\end{equation}
Here, $V_{Dirac}^{\pm}$ are related to $W(r)$ as,
\begin{equation} \label{potential}
V_{Dirac}^{\pm} = \mp f \left( \frac{dW}{dr} \right) + W^2
\end{equation}
In a paper by Anderson and Price \cite{ander} it was discussed that two potentials $V_{Dirac}^{+}$ and $V_{Dirac}^{-}$ as given above which are related, will produce the same physical consequences. Hence, both potentials  will produce the same  QNM spectra. Therefore, we will use only $V_{Dirac}^+$ for all of our computations in the rest of the paper and will be referred to as just $V_{Dirac}(r)$.

\begin{figure} [H]
\begin{center}
\includegraphics{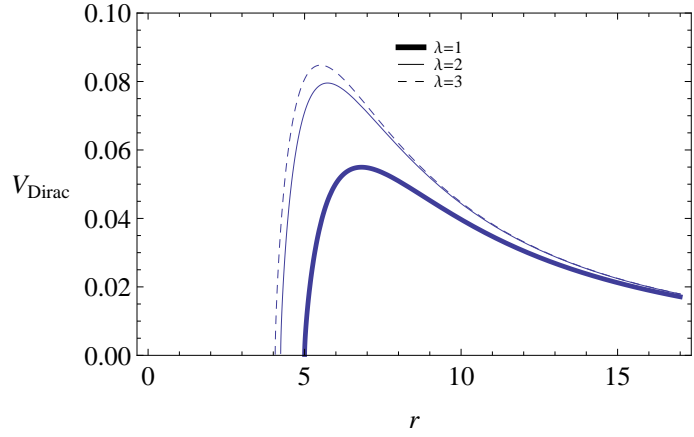}
\caption{The figure shows  $V_{Dirac}(r)$ vs $r$. Here $M = 2, \mu =3$ and $Q = 1$.}
\label{potlambda}
 \end{center}
 \end{figure}

\begin{figure} [H]
\begin{center}
\includegraphics{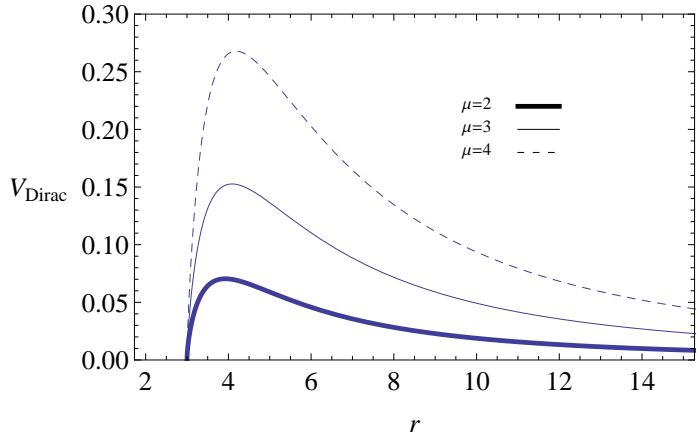}
\caption{The figure shows  $V_{Dirac}(r)$ vs $r$. Here $M = 1, \lambda =1$ and $Q = 1$.}
\label{potmu}
 \end{center}
 \end{figure}

\begin{figure} [H]
\begin{center}
\includegraphics{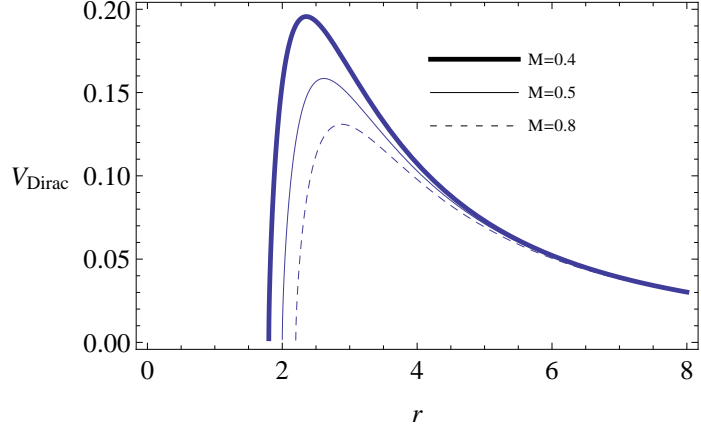}
\caption{The figure shows  $V_{Dirac}(r)$ vs $r$. Here $\lambda =1, \mu =2$ and $Q = 1$.}
\label{potmass}
 \end{center}
 \end{figure}

\begin{figure} [H]
\begin{center}
\includegraphics{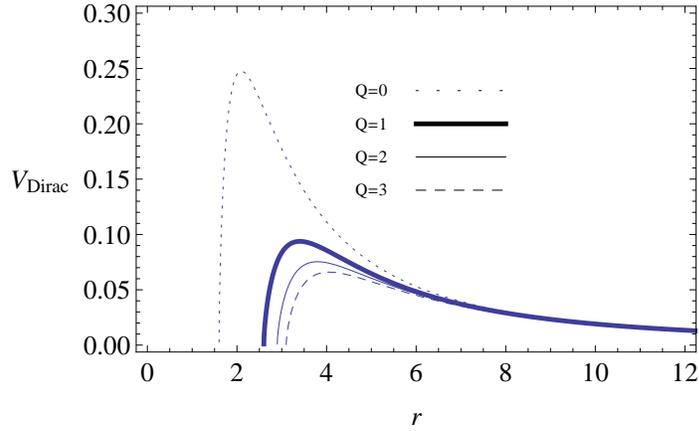}
\caption{The figure shows  $V_{Dirac}(r)$ vs $r$. Here $M = 0.8, \lambda = 1$ and $\mu = 2$.}
\label{potq}
 \end{center}
 \end{figure}

The effective potential $V_{Dirac}(r)$ depends on four parameters: $M, Q, \mu$ and $\lambda$. In Fig.$\refb{potlambda}$, $V_{Dirac}(r)$ is plotted as a function of $r$ by varying $\lambda$. When $\lambda$ increases, the height of the potential increases. In Fig.$\refb{potmu}$, $V_{Dirac}(r)$ is plotted as a function of $r$ by varying $\mu$. When $\mu$ increases, the height of the potential increases. In Fig.$\refb{potmass}$  $V_{Dirac}(r)$ is plotted as a function of $r$ by varying $M$. When mass increases, the height of the potential decreases. In Fig.$\refb{potq}$ , $V_{Dirac}(r)$ is plotted as a function of $r$ by varying $Q$. When $Q$ increases, the potential height decreases. However, the potential for the Schwarzschild black hole (with $Q=0$) is higher than any one of them. The behavior of the potential $V_{Dirac}$ is similar to the behavior of the effective potential of the massless scalar field around the massive gravity black hole \cite{fernando8}.

%%%%%%%%%%%%%%%%%%

\section{ Computation of  QNM frequencies }

QNM of a black hole for a particular perturbation are given by the solutions to the corresponding wave equations. In this paper, the QNM are given by the solution to the Dirac wave equation in eq$\refb{final}$. To seek solutions boundary conditions has to be imposed; for asymptotically flat black hole, such as the one considered in this paper,  the boundary conditions are, ingoing waves at the horizon and out going waves at the spatial infinity. In this work we are particularly interested in frequencies of QNM. QNM frequencies are complex and do not form a complete set. For stable black holes, the imaginary part of the frequency $\omega$ is expected to be negative. The modes of the perturbations are labeled by an integer $n$. The fundamental mode is described by $n =0$ mode. Since the largest value of $\omega_I$ is for $ n=0$, we will  mostly be interested on $\omega_I$ with $n=0$.

Analytical solutions to the wave equations resulting from black hole perturbations are rare. There are few examples we would like to mention here. QNM frequencies of dilaton black holes in 2+1 dimensions have been found by Fernando in \cite{fernando1} \cite{fernando2} \cite{fernando3}. Some other examples with exact QNM frequencies are in 
 \cite{kwon} \cite{bertha} \cite{yun}.

Due to the inability to solve the wave equations and frequencies analytically, there are numerous numerical methods developed to compute QNM frequencies.   The reader is referred to the  review by Konoplya  and Zhidenko \cite{kono1} as an excellent source of reference for various methods. Out of the methods used in literature, the WKB analysis is one of the most popular methods to find QNM frequencies. WKB approach was developed by Iyer and Will  \cite{will} and subsequently developed to sixth order by Konoplya \cite{kono4}. This method has been employed to compute QNM frequencies corresponding to  black hole perturbations in several cases. Examples are found in  \cite{fernando6}\cite{fernando7}\cite{fernando8}. In this method,  the QNM frequencies are given by the expression,

\be
\omega^2 = - i \sqrt{ - 2 V''(r_{max})} \left( \Sigma^6_{i=2}  \Omega_i  +   n + \frac{ 1}{2} \right) + V(r_{max})
\ee
Here, $r_{max}$ is the place where $V(r)$ is maximum and $V''(r)$ is the second derivative of the potential.  Expressions for $\Omega_i$ can be found in \cite{kono4}. 

In  this paper, we used the WKB approximation described above to compute the QNM frequencies. We used all 6 orders for the computation. WKB approximation is accurate when the multipole number $\mu$ is greater than the overtone number $n$ \cite{kono4}.

We have computed $\omega$ for the Dirac field by varying the parameters in the theory, $M, \lambda, Q, n$ and $\mu$. All $\omega_I$ values were negative; therefore, the black hole is stable under Dirac field  perturbations.  We have presented results in tables as well as figures. Notice that $\omega_I$ is negative and we have only written the value without the negative sign in the table. We have also used only the value in plotting the graphs.

\begin{figure} [H]
\begin{center}
\includegraphics{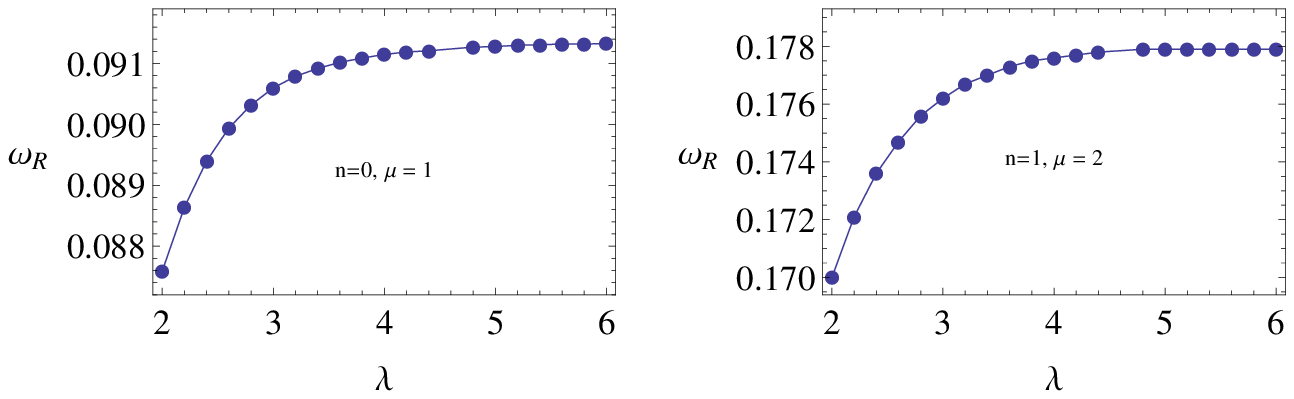}
\caption{The figure shows  $\omega_R$ vs $\lambda$. Here $M = 2, Q = 1$ and $\mu =1,2$.}
\label{lambreal}
 \end{center}
 \end{figure}

\begin{figure} [H]
\begin{center}
\includegraphics{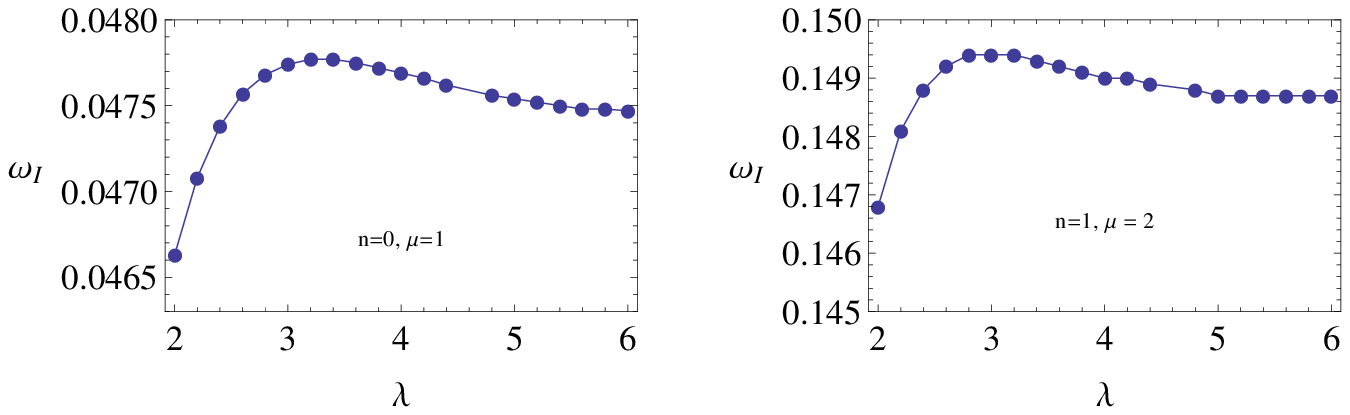}
\caption{The figure shows  $\omega_I$ vs $\lambda$. Here $M = 2, Q = 1$ and $\mu =1,2$.}
\label{lambima}
 \end{center}
 \end{figure}

First, let us discuss the effects of $\lambda$ on $\omega$. The values are given in Table 1. For small values of $\lambda$, both $\omega_R$ and $\omega_I$ increases with $\lambda$. When $\lambda$ is increased further, $\omega_R$ approaches a constant value. The behavior is similar for both $n=0$ and $n=1$, and  is presented in Fig.$\refb{lambreal}$. For $\omega_I$, increasing $\lambda$ further increase it to a maximum and then decreases to a constant value. This behavior is demonstrated in Fig.$\refb{lambima}$. The stable values reached by both $\omega_R$ and $\omega_I$ corresponds to the values for the Schwarzschild black hole given by, 
$\omega_{Sch} =  0.09133- i 0.04747$($\mu =1$) and $0.1779 -i 0.1487$($ \mu =2$). In general,  the Dirac field decay slower around the massive gravity black hole for small $\lambda$. There is a range of $\lambda$ for which the Dirac field decay faster than the Schwarzschild black hole. Hence the significance of $\lambda$ on the decay rates are only for a short range  as is obvious from the graphs. This behavior is not surprising since the metric function $f(r) \approx 1 - \frac{ 2 M}{r}$ for  large $\lambda$ values. The behavior of the scalar field around this black hole was similar as shown in \cite{fernando8}. 

\begin{figure} [H]
\begin{center}
\includegraphics{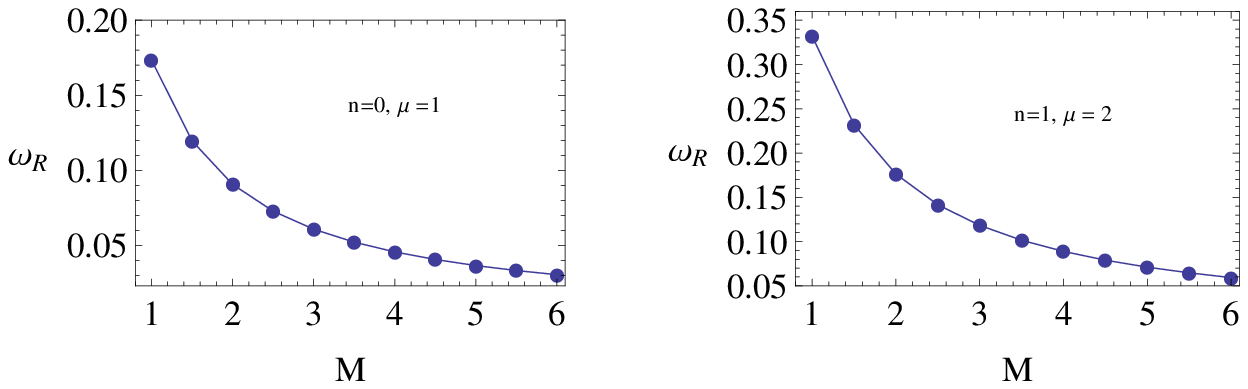}
\caption{The figure shows  $\omega_R$ vs $M$. Here $\lambda = 3, Q = 1$ and $\mu =1,2$.}
\label{massreal}
 \end{center}
 \end{figure}

\begin{figure} [H]
\begin{center}
\includegraphics{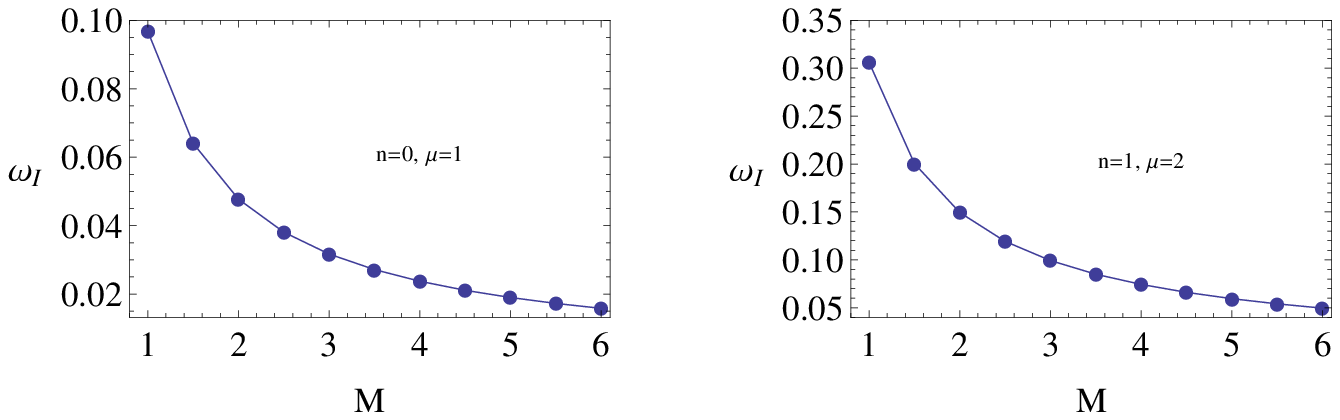}
\caption{The figure shows  $\omega_I$ vs $M$. Here $\lambda = 3,  Q=1$ and $\mu =1,2$.}
\label{massima}
 \end{center}
 \end{figure}

When $\omega_R$ and $\omega_I$ were studied by varying the mass of the black hole as given in Table 2, both decreases with increasing mass $M$ as shown in Fig.$\refb{massreal}$ and Fig.$\refb{massima}$. Hence, small black holes are more stable compared to large black holes.  The scalar field behaves similarly \cite{fernando8}.

\begin{figure} [H]
\begin{center}
\includegraphics{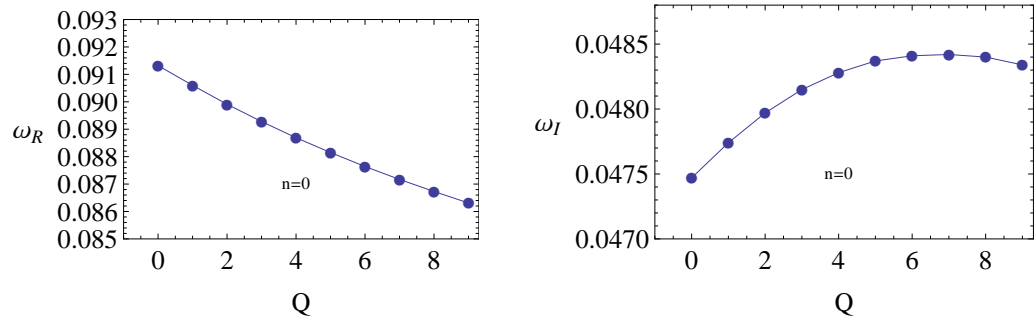}
\caption{The figure shows  $\omega_R$ and $\omega_I$  vs $Q$. Here $M = 2, \lambda = 3$ and $\mu =1$.}
\label{chargefre}
 \end{center}
 \end{figure}

When $Q$ parameter is varied,   $\omega_R$ decreases  with $Q$.  On the other hand  $\omega_I$  increases to a mamximum and then decreases  as shown in Table 3 and Fig.$\refb{chargefre}$.  Hence there is a maximum value of  $Q$ which gives more stability.  Interestingly, the scalar field demonstrated different  behavior by increasing  $\omega_I$ when $Q$ is increased \cite{fernando8}.

\begin{figure} [H]
\begin{center}
\includegraphics{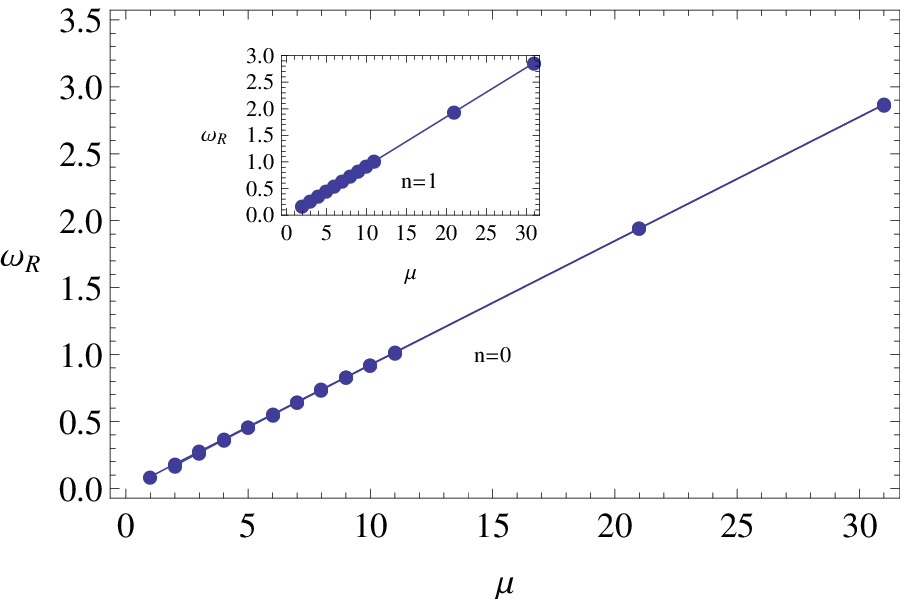}
\caption{The figure shows  $\omega_R$ vs $\mu$. Here $M = 2, Q = 1$ and $\lambda =2$.}
\label{sphereal}
 \end{center}
 \end{figure}

\begin{figure} [H]
\begin{center}
\includegraphics{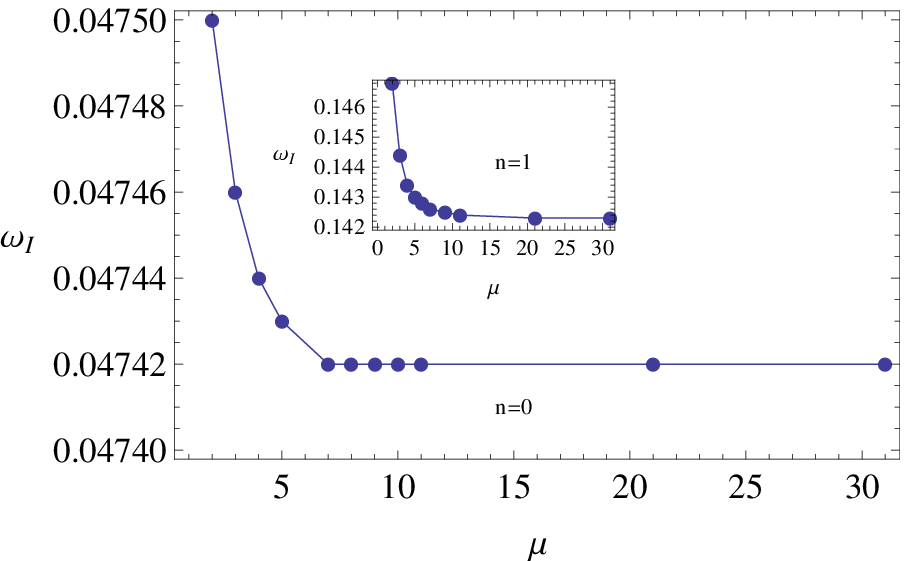}
\caption{The figure shows  $\omega_I$ vs $\mu$. Here $M = 2, Q = 1$ and $\lambda =2$.}
\label{spheima}
 \end{center}
 \end{figure}

In Fig.$\refb{sphereal}$ and Fig.$\refb{spheima}$, $\omega_R$ and $\omega_I$ are plotted against the multipole number  $\mu$.  The values are given in Table 4. When $\mu$ is increased, $\omega_R$ increases linearly. For both $n=0$ and $n=1$, this behavior persists. On the other hand, $\omega_I$ decreases to a constant value when $l$ is increased. The behavior is similar for $n=0$ and $n=1$.

\begin{figure} [H]
\begin{center}
\includegraphics{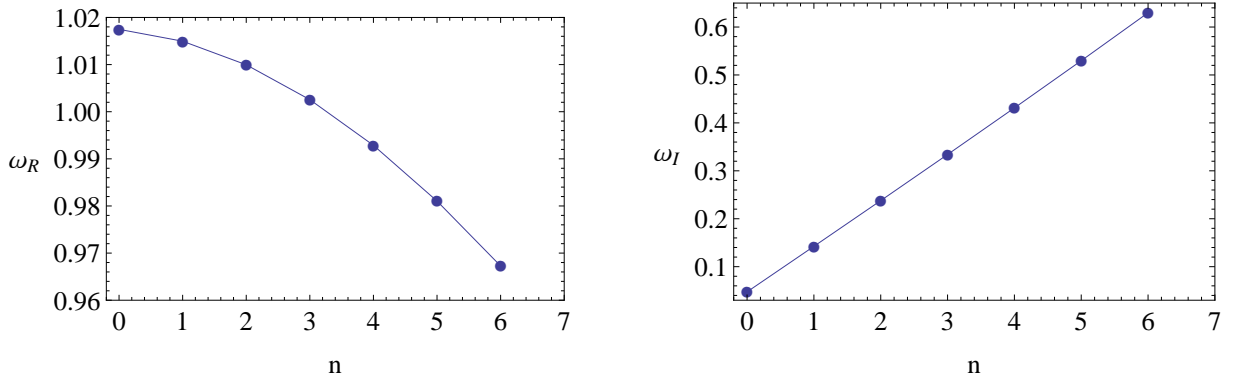}
\caption{The figure shows  $\omega_R$ and $\omega_I$ vs $n$. Here $M = 2, Q = 1, \mu=11$ and $\lambda =2$.}
\label{freforn}
 \end{center}
 \end{figure}

When $\omega_R$ and $\omega_I$ is plotted by changing the mode number $n$, $\omega_R$ decreases  with $n$ and $\omega_I$ increases with $n$.  Also, $\omega_I$ has a linear relation with $n$. This is demonstrated in Fig.$\refb{freforn}$ and Table 5.

It was observed that the temperature of the black hole depend on  $\omega_I$  linearly for asymptotically anti de Sitter black holes $\cite{horo11}$. Hence, 
we also plotted $\omega_I$ vs the temperature of the black hole by varying  $Q$. A linear relation seems to exist between the temperature and $\omega_I$

%%%%%%%%%%%%table for lambda %%%%%%%%%%%

\begin{center}
\begin{tabular}{|l|l|l|l|l|r} \hline \hline
 $\lambda$  & $\omega_R$ (n =0, $\mu =1$)  &  $\omega_I$( n =0, $\mu =1$) & $\omega_R$ ( n =1, $\mu =2$) & $ \omega_I$ ( n =1, $\mu =2$) \\ \hline

2 &  0.08759 & 0.04663 & 0.1700 & 0.1468  \\ \hline
2.2 &  0.08864 & 0.04708 & 0.1721 & 0.1481    \\ \hline
2.4 & 0.08940 & 0.04738 & 0.1736 & 0.1488   \\ \hline
2.6 &  0.08994 & 0.04757 & 0.1747 & 0.1492   \\ \hline
2.8 & 0.09032 & 0.04768 & 0.1756 & 0.1494    \\ \hline
3 & 0.09060 & 0.04774 & 0.1762 & 0.1494   \\ \hline
3.2 & 0.09079 & 0.04777 & 0.1767 & 0.1494      \\ \hline
3.4 & 0.09092 & 0.04777 & 0.1770 & 0.1493   \\   \hline
3.6 &   0.09102 & 0.04775 & 0.1773 & 0.1492  \\ \hline
3.8 &  0.09109 & 0.04772 & 0.1775 & 0.1491    \\ \hline
4 &  0.09115 & 0.04769 & 0.1776 & 0.1490   \\ \hline
4.2 &  0.09119 & 0.04766 & 0.1777 & 0.1490 \\ \hline
4.4 &  0.09121 & 0.04762 & 0.1778 & 0.1489  \\ \hline
4.6 & 0.09124 & 0.04759 & 0.1778 & 0.1488 \\ \hline
4.8 &   0.09127 & 0.04756 & 0.1779 & 0.1488 \\ \hline
5.0 &   0.09128 & 0.04754 & 0.1779 & 0.1487 \\ \hline
5.2 &    0.09130 & 0.04752 & 0.1779 & 0.1487\\ \hline
5.4 &   0.09131 & 0.04750 & 0.1779 & 0.1487 \\ \hline
5.6 &    0.09132 & 0.04748 & 0.1779 & 0.1487\\ \hline
5.8 &    0.09132 & 0.04747 & 0.1779 & 0.1487\\ \hline
6.0 &   0.09133 & 0.04747 & 0.1779 & 0.1487 \\ \hline

\end{tabular}
\end{center}

Table 1: QNM frequencies for various $\lambda$ values. Here $ M = 2, Q =1$.

%%%%%%%%%%%%table for M %%%%%%%%%%

\begin{center}
\begin{tabular}{|l|l|l|l|l|r} \hline \hline
 M  & $\omega_R$ (n =0, $\mu =1$)  &  $\omega_I$( n =0, $\mu =1$) & $\omega_R$ ( n =1, $\mu =2$) & $ \omega_I$ ( n =1, $\mu =2$) \\ \hline

1 &  0.1735 & 0.09679 & 0.3321 & 0.3061  \\ \hline
1.5  & 0.1196 & 0.06406 & 0.2320 & 0.2005  \\ \hline
2 &   0.09060 & 0.04774 & 0.1762 & 0.1494  \\ \hline
2.5 &  0.07276 & 0.03809 & 0.1416 & 0.1193   \\ \hline
3 &   0.06074 & 0.03170 & 0.1183 & 0.0993   \\ \hline
3.5 &  0.05211 & 0.02716 & 0.1015 & 0.0850   \\ \hline
4 &      0.04562 & 0.02375 & 0.0889 & 0.0744  \\ \hline
4.5 &   0.04056 & 0.02111 & 0.0790 & 0.0661 \\   \hline
5 &    0.03651 & 0.01900 & 0.0711 & 0.0595  \\ \hline
5.5 &   0.03320 & 0.01727 & 0.0647 & 0.0541    \\ \hline
6 &  0.03043 & 0.01583 & 0.0593 & 0.0496  \\ \hline

\end{tabular}
\end{center}

 Table 2: QNM frequencies for various $M$ values. Here $ \lambda = 3, Q =1$.
 
 %%%%%%%%table for Q%%%%%%%%
 \begin{center}
\begin{tabular}{|l|l|l|r|} \hline \hline 
 Q & $\omega_R$  &  $\omega_I$ & Temperature \\ \hline

0 &  0.09132   &  0.04747  &  0.0198944  \\ \hline
1 & 0.09060 & 0.04774 & 0.0201830  \\ \hline
2 & 0.08991 &  0.04797 & 0.0204320  \\ \hline
3 & 0.08928 & 0.04815 & 0.0206480  \\ \hline
4 & 0.08869 & 0.04828 & 0.0208380  \\ \hline
5 & 0.08815 & 0.04837 & 0.0210050  \\ \hline
6 & 0.08764 & 0.04841 & 0.0211536  \\ \hline
7 & 0.08717 & 0.04842 & 0.0212854  \\ \hline
\end{tabular}
\end{center}
 
 Table 3: QNM frequencies for Q values. Here $ M = 2, \lambda =3$ and $ \mu =1$.
 
 %%%%%%%%%%%table for mu %%%%%%
 
\begin{center}
\begin{tabular}{|l|l|l|l|l|r} \hline \hline
 $\mu$  & $\omega_R$ (n =0)  &  $\omega_I$( n =0) & $\omega_R$ ( n =1) & $ \omega_I$ ( n =1) \\ \hline

2 &  0.1826 & 0.04750  & 0.1700 & 0.1468    \\ \hline
3 &  0.2760 & 0.04746 & 0.2671 & 0.1444   \\ \hline
4 &  0.3689 & 0.04744 & 0.3621 & 0.1434  \\ \hline
5 & 0.4617 & 0.04743 & 0.4562 & 0.1430    \\ \hline
6 & 0.5544 & 0.04743 & 0.5498 & 0.1426   \\ \hline
7 & 0.6471 & 0.04742 & 0.6431 & 0.1426     \\ \hline
8 & 0.7397 & 0.04742 & 0.7362 & 0.1426   \\   \hline
9 &   0.8323 & 0.04742 & 0.8292 & 0.1425  \\ \hline
10 & 0.9249 & 0.04742 & 0.9221 & 0.1424    \\ \hline
11 & 1.0175 & 0.04742 & 1.0150 & 0.1423   \\ \hline
21 &  1.9431 & 0.04742 & 1.9418 & 0.1423 \\ \hline
31 &  2.8686 & 0.04742 & 2.8677 & 0.1423  \\ \hline

\end{tabular}
\end{center}

Table 4: QNM frequencies for various $\mu$ values. Here $ M = 2, \lambda =2$ and $Q =1$.

%%%%%%%%%%%%table n%%%%%%%%%%%%%%

\begin{center}
\begin{tabular}{|l|l|l|r|} \hline \hline 
 n & $\omega_R$  &  $\omega_I$  \\ \hline

0 &  1.01748 & 0.04742 \\ \hline
1 &   1.01496 & 0.14241\\ \hline
2 &   1.00996 & 0.23784 \\ \hline
3 &   1.00256 & 0.033402 \\ \hline
4 &  0.099287 & 0.43122  \\ \hline
5 &   0.98104 & 0.52972 \\ \hline
6 &   0.96727 & 0.62975 \\ \hline

\end{tabular}
\end{center}

 Table 5: QNM frequencies for various overtone mode number $n$ values. Here $ M = 2, \lambda =2, \mu =11$ and $Q =1$.

%%%%%%%Figures%%%%%%%%%%%%%%%%%%%%%%

 \begin{figure} [H]
\begin{center}
\includegraphics{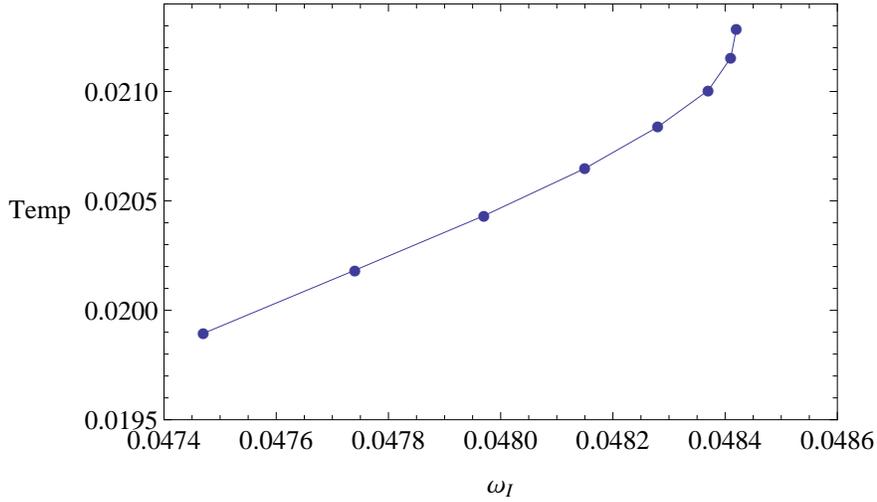}
\caption{The figure shows  $Temp$ vs $\omega_I$. Here $M = 2, \mu=1$ and $\lambda =3$. Here $Q$ is varied}
\label{tempomega}
 \end{center}
 \end{figure}

%%%%%%%%%%%%%%%%%%%%%%%%%%%%%%%%%%%%%%%

\section{ Analytical values of $\omega$ with P$\ddot{o}$schl-Teller approximation}

One can obtain analytical values of $\omega$ by approximating the effective potential with the well known P$\ddot{o}$schl-Teller potential.  Ferrari and Mashhoon \cite{mash}, applied this method to obtain exact results for the Schwarzschild black hole, Reissner-Nordstrom black hole and the Kerr black hole. Since the wave equation with the P$\ddot{o}$schl-Teller potential can be solved exactly, one can find  analytical formulas for the QNM frequencies $\omega$. Here, we will apply this method to obtain approximate values  for QNM frequencies of the Dirac field with  large $\lambda$.

In  the  the P$\ddot{o}$schl-Teller approximation, the effective potential is approximated by,
\be
V = \frac{ V_o}{ Cosh^2 \beta(r_* - r_{*o}) }
\ee
Here $r_*$ is the tortoise coordinate described in eq.$\refb{tortoise}$.   $r_{*o}$ is the point where the potential has its  the maximum:  hence $\frac{dV}{dr_*} = 0$ at $r_* = r_{*o}$. The variable $\beta$  and $V_0$ are given by,
\be
\beta^2 = \frac{1}{ 2 V_o} \frac{d^2V}{ dr_*^2} |_{r = r_{*o}}
\ee
\be
V_o = V(r_* = r_{*o})
\ee
It was shown in \cite{cardoso2} \cite{mash} that the QNM frequencies $\omega$  for the above potential is given by,
\be
\omega = \pm \sqrt{ V_o - \beta^2/4} - i \beta ( n + 1/2)
\ee
Due to the complicated nature of the potential in eq.$\refb{potential}$, we will study the QNM frequencies at the eikonal limit (large $\mu$). Then, the dominant term in the effective potential is the one proportional to $\mu^2$. Hence,
\be
V(r) \approx \frac{ \mu^2 f}{r^2}
\ee
Since we are studying the wave equation for large $\lambda$,  the function $f(r)$ can be approximated  as,
\be \label{poshf}
f(r) \approx  1 - \frac{ 2 M}{r} - \bar{Q} \kappa
\ee
Here $\kappa = \frac{1}{\lambda}$ and $\bar{Q}$ is chosen such that  $f$ will be dimensionally correct. When $\lambda \ra \infty$, $f \approx 1 - \frac{ 2 M}{r}$. Hence for large $\lambda$, the horizon gets closer to the horizon of the Schwarzschild black hole which is what is expected as given  in  Fig.$\refb{frlambda}$.

Now, $\frac{dV}{dr_*} = \left(\frac{dV}{ dr}\right) \left( \frac{ dr} {dr_*}\right) = 0$ lead to the solutions,
\be
r_o = \frac{ 3 M}{ (1 - \bar{Q} \kappa)} 
\ee
The final results for $\beta$ and $V_o$ are given as,
\be
\beta^2 = \frac{ ( 1 - \bar{Q}\kappa)^4} { 27 M^2}
\ee
\be
V_o = \mu^2 \frac{ ( 1 - \bar{Q} \kappa)^3 }{ 27 M^2}
\ee
Hence $\omega$ is given as,
\be \label{posh}
\omega = \frac{ ( 1 - \bar{Q}\kappa)^2}{ 3 \sqrt{3} M} \sqrt{ \frac{\mu^2}{( 1- \bar{Q} \kappa)} - \frac{1}{4}} -  i \frac{ ( 1 - \bar{Q}\kappa)^2}{ 3 \sqrt{3} M}
\ee
It is clear that when $\lambda \ra \infty ( \kappa \ra 0)$, $\omega$ reaches the value for the Schwarzschild black hole obtained by Ferrari and Mashhoon in \cite{mash}.

By observing the $\omega$ given above, one can explain the behavior of $\omega$ obtained with the WKB approach.  For example, when $\mu$ becomes large, the above approximation gives $\omega_R \propto \mu$  which was observed in Fig.$\refb{sphereal}$.  On the other hand $\omega_I$ is independent of $\mu$ which is what was demonstrated in Fig.$\refb{spheima}$. When $\kappa \ra 0 ( \lambda \ra \infty)$,  $\omega \ra \omega_{Schwarzschild}$. This is clear from   Fig.$\refb{lambreal}$ and Fig.$\refb{lambima}$. When the mass $M$ is increased, both $\omega_R$ and $\omega_I$ decreases from eq.$\refb{posh}$. This is the same behavior observed in Fig.$\refb{massreal}$ and Fig.$\refb{massima}$.

%%%%%%%%%%%%%%%%%%%

\section{ Conclusion}

We have studied QNM frequencies of a spin $\frac{1}{2}$, massless Dirac field perturbations of a black hole in massive gravity. First we computed frequencies $\omega$ with the 6th order WKB approximation. The parameters in the theory, $\lambda, M, Q$ and $\mu$ are varied to  obtain the relations with $\omega$. For small values of $\lambda$, both $\omega_R$ and $\omega_I$ increases. When $\lambda$ gets larger, $\omega_R$ reaches a stable value;  $\omega_I$ keep increasing to a maximum and then decreases to a stable value. The stable value corresponds to the $\omega$ of the Schwarzschild black hole. Similar behavior was observed for the massless scalar field perturbations around the black hole \cite{fernando8}.

When the mass $M$ is increased, both $\omega_R$ and $\omega_I$ decreases. Hence, small black holes are  more stable. When $Q$ increases, $\omega_R$ decreased. On the other hand $\omega_I$ increased to a maximum before falling off for large Q. Hence there is a maximum value of Q for which the black hole is most stable.

When the multipole number $\mu$ is increased, $\omega_R$ increase linearly; $\omega_I$  decreases and reach a stable value. This behavior is consistent for the mode values $n=0$ and $n=1$. When the mode number $n$ is increased, $\omega_R$ decreases; $\omega_I$ increases linearly with $n$.

We used P$\ddot{o}$schl-Teller approximation to find an analytical values  for  $\omega$  when $\lambda$ is large. Our computation was done for large $\mu$.  The analytical $\omega$ obtained in this method verified the numerical values obtained with the WKB approach.

As future work, it would be interesting to find the stability under electromagnetic perturbations. Also, the cross sections of fields of spin $0, \frac{1}{2}, 1$ would be an interesting aspect to study.

\vspace{0.5 cm}

%%%%%%%%%%%%%%%%%%%%%

{\bf Acknowledgments:}  The author wish to thank R. A. Konoplya for providing the  {\it Mathematica}  file for  WKB approximation.

%%%%%%%%%%%%%%%%%%%%%%

\end{document}